\documentclass[12pt]{article}
\begin{document}
\pagestyle{plain}
\title{The M\"ossbauer effect in homogeneous magnetic field}
\author{Miroslav Pardy\\
Department of Physical Electronics \\
Masaryk University \\
Kotl\'{a}\v{r}sk\'{a} 2, 611 37 Brno, Czech Republic\\
e-mail:pamir@physics.muni.cz}
\date{\today}
\maketitle
\vspace{50mm}

\begin{abstract}
We derive the probability of the M\"ossbauer effect
realized  by the charged particle moving in the homogeneous magnetic field, or, in accelerating field. The submitted approach represents new deal of the M\"ossbauer physics.  
\end{abstract}

\vspace{1cm}


\section{Introduction}

The discovery of the M{\"o}ssbauer effect is  unique.  
The emission and absorption of x-rays by gases had been observed previously, and it was expected that the resonance effects  would be found for gamma rays, which are created by nuclear transitions (as opposed to x-rays, which are typically produced by electron transitions). However, attempts to observe nuclear resonance produced by gamma-rays in gases failed due to recoil, preventing resonance (the Doppler effect also broadens the gamma-ray spectrum). M{\"o}ssbauer  was able to observe resonance in nuclei of solid iridium, as opposite to no gamma-ray resonance  in gases. He proposed that, for the case
of atoms bound into a solid, the nuclear events could occur essentially without the recoil.

The photon emission energy for the atom of iridium  is approximately ${\rm 1\;eV}$ and the recoil energy is  $E_{recoil}(optical) = \;$  GeV = $10^{-11}\;$ eV . (Rohlf, 1994).
On the other hand the $\gamma$-emission energy of the iridium nucleus is 
$\approx {\rm 10^{5}\; eV}$. Then, the recoil energy caused by the emission of such $\gamma$-ray is $E_{recoil}  \approx 10^{-1}\;$ eV (Rohlf, 1994). So, we see that the recoil caused by the $\gamma$-ray emission is substantially greater  than the recoil caused by the optical emission and we can expect the big shift of spectral lines in the nuclear system.  

The motion of the decaying excited nucleus of iridium $^{191}{\rm Ir}^{*}$ causes the Doppler broadening and the Doppler shift of the gamma spectrum. Let us consider the motion of the excited iridium in the direction of the emitted photon. The Doppler formula for the Lorentz boosted proton energy $E'$ is as follows (Rohlf, 1994):

$$E' = E\frac{\sqrt{1 + v/c}}{1 - v/c} = E\gamma(1 - v/c)
\approx E(1 + v/c); .\eqno(1)$$

The fractional change of he proton energy is $(E'- E)/E = v/c. $ 

The resonance (the overlapping of emission and absorption curves) is destroyed if $v/c$ is equal few times $\Gamma/E = (\hbar/\tau)/E ( = 2.7 \times 10^{-11}$ for iridium), where $\Gamma$ is the natural spectral line width and $\tau$ is the life time of the excited state of nucleus. For $v/c = \Gamma/E$, we get $v/c = {\rm 2.7\times 10^{-11}}$. The corresponding speed is $v = {\rm (3\times 10^{8}\; m/s)\times (2.7 \times 10^{-11}) \approx 10^{-2}\; m.s^{-1}}$.  So, the speed of centimeters per second destroys the resonance absorption. In other words, the overlap of the very narrow absorption and emission curves is zero for the nuclear system.

In a solid, the nuclei are bound to the lattice and do not recoil in the same way as in a gas. The lattice as a whole with mass $M$ recoils but the recoil energy is negligible because $M$  is the mass of the whole lattice. However, the energy in a decay can be taken up or supplied by lattice vibrations. The energy of these vibrations is quantized in units known as phonons. The M\"ossbauer effect occurs because there is a finite probability of a decay occurring involving no phonons. Thus, the entire crystal acts as the recoiling body, and these events are essentially recoilless. In these cases, since the recoil energy is negligible, the emitted gamma rays have the appropriate energy and resonance can occur.

Gamma rays have very narrow line widths. This means they are very sensitive to small changes in the energies of nuclear transitions. In fact, gamma rays can be used as a probe to observe the effects of interactions between a nucleus and its electrons and those of its neighbors. This is the basis for M{\"o}ssbauer spectroscopy, which combines the M{\"o}ssbauer effect with the Doppler effect to monitor such interactions.
  
\section{The quantum theory of the M{\"o}ssbauer effect in the homogeneous magnetic field}

We can define the M\"ossbauer effect in homogeneous magnetic field  as the analogue of the M\"ossbauer effect for crystal,
where particle in crystal is replaced by the charged particle in homogeneous magnetic field.
We consider the situation where the nucleus emitting the gamma rays is inbuilt (implanted) in homogeneous magnetic field.
The initial state of the crystal  let be $\psi_{crystal}$ and the final state of crystal let be $\psi_{crystal}$. Then, according to Feynman (1972),  there is a probability of no recoil after the  photon emission with momentum $\bf k$ from nucleus. The amplitude of probability is
$({\bf k} = {\bf p}/\hbar)$
$$a = \left<\psi_{crystal}|e^{i({\bf k}\cdot{\bf r})}|\psi_{crystal}\right> ,\eqno(2)$$
here ${\bf r}$ is the displacement of lattice atom. 

The probability of unchanging of the basic magnetic state $\psi_{0}$ (the analogue of the persistence of vacuum in quantum field theory) after the $\gamma$-emission is  then  
 $P = a^{2}$. So,  we have

$$P =   \left|\left<\psi_{0}\left|e^{i{\bf k}\cdot{\bf r}}\right|\psi_{0}\right>\right|^{2} =
 \left|\int e^{i{\bf k}\cdot{\bf r}}|\psi_{0}|^{2}d{\bf r}\right|^{2}, \eqno(3)$$
where the exponential function in (3) can be expanded using the partial amplitudes taken from the textbooks of quantum mechanics of scattering processes as follows:

$$e^{i{\bf k}\cdot{\bf r}} = 4\pi\sum_{l=0}^{\infty}\sum_{m = -l}^{l}i^{l}j_{kr}Y_{lm}(\Theta,\Phi)
Y_{lm}^{*}(\theta,\phi).  \eqno(4)$$

The mathematical term  $i{\bf k}\cdot{\bf r}$ can be written using the azimuthal angle $\varphi$ for the process in the plane of motion in the magnetic field as $ikr\cos\varphi$.
So, we shall calculate the probability corresponding to the situation where the crystal  is replaced by the homogeneous magnetic field.
 
We take the basic function $\psi_{0}$ for one
electron in the lowest Landau level, as 

$$\psi_{0} = \left(\frac{m\omega_{c}}{2\pi\hbar}\right)^{1/2}
\exp\left(-\frac{m\omega_{c}}{4\hbar}(x^{2} + y^{2})\right), \eqno(5)$$ 
which is solution of the Schr\"odinger equation in the magnetic field with potentials ${\bf A} = (-Hy/2, Hx/2, 0)$,  $A_{0} = 0$
(Drukarev, 1988):

$$\left[\frac{p_{x}^{2}}{2m} + \frac{p_{y}^{2}}{2m} - \frac{m}{2}
\left(\frac{\omega_{c}}{2}\right)^{2}(x^{2} + y^{2})\right]\psi = E\psi.
\eqno(6)$$ 

So, The main problem is to calculate the integral in the polar coordinates $r, \varphi$ as follows:

$$I = \int_{0}^{2\pi}\int_{0}^{\infty}d\varphi rdr |\psi_{0}|^{2}e^{ikr\cos\varphi},
 \eqno(7)$$
which can be simplified introducing constants $C$ and $\alpha$ as follows ($Q$ is a charge of the  M\"ossbauer particle,
$c$ is the velocity of light): 

$$C = \left(\frac{m\omega_{c}}{2\pi\hbar}\right)^{1/2}; \quad \alpha = \left(\frac{m\omega_{c}}{4\hbar}\right); \quad \omega _{c} = \frac{|Q|H}{mc}.\eqno(8)$$

Then, 

$$I = C^{2}\int_{0}^{2\pi}\int_{0}^{\infty}d\varphi rdr e^{-2\alpha r^{2}}e^{ikr\cos\varphi}.
 \eqno(9)$$
Let us firs consider the calculation of the polar integral of the form:

$$I_{1} = \int_{0}^{2\pi}[\cos(kr\cos\varphi) + i\sin(kr\cos\varphi)]d\varphi.\eqno(10)$$

Using identities

$$\cos(a\cos\varphi) = J_{0}(a) + 2\sum_{n = 1}^{\infty}(-1)^{n}J_{2n}(a)\cos(2n\varphi),
\eqno(11)$$

$$\sin(a\cos\varphi) = 2\sum_{n = 1}^{\infty}(-1)^{n + 1}J_{2n - 1}(a)\cos[(2n-1)\varphi)],
\eqno(12)$$
where $J_{n}$ are the Bessel functions, we get after integration that 

$$I_{1} = J_{0}(kr),\eqno(13)$$
where the Bessel function $J_{0}$ can be expressed as the series 

$$J_{0}(x)  =  \sum_{k = 0}^{\infty} \frac{(-1)^{k}x^{2k}}{2^{2}2^{4}....(2k)^{2}} = 
1 - \frac{x^{2}}{2^{2}} + \frac{x^{4}}{2^{2}4^{2}} - \frac{x^{6}}{2^{2}4^{2}6^{2}} + ...
\eqno(14)$$

So, The following step is, to calculate the following integral:

$$I_{2} = \int_{0}^{\infty}J_{0}(kr)e^{-2\alpha r^{2}} rdr\eqno(15)$$

If we restrict the calculation with the approximate Bessel function, then we get for the probability of the persistence of the state in the form:

$$P \approx C^{2}\left|2\pi\int_{0}^{\infty}rdr\left[e^{-2\alpha r^{2}} - e^{-2\alpha r^{2}}\frac{(kr)^{2}}{2^{2}}\right]\right|^{2} \eqno(16)$$

Using the integrals 

$$\int_{0}^{\infty} e^{-2\alpha r^{2}} rdr = \frac{1}{4\alpha}; \quad \eqno(17)$$

$$\int_{0}^{\infty} \frac{k^{2}r^{3}}{2^{2}}e^{-2\alpha r^{2}} dr = 
\frac{k^{2}}{32}\frac{1}{\alpha^{2}},
\eqno(18)$$
where the integrals are the special cases of the table integral (Gradshteyn and Ryzhik, 2007a)

$$\int_{0}^{\infty}x^{2n+1} e^{-p x^{2}} dx = \frac{n!}{2p^{n + 1}}; \quad p>0,\eqno(19)$$
we get the final approximation formula for the existence of the M\"ossbauer effect in magnetic field realized by the decay of the charged ion. Or, 

$$ P \approx  4\pi^{2}C^{2}\left|\frac{1}{4\alpha} + \frac{k^{2}}{32\alpha^{2}}\right|^{2}
\eqno(20)$$

Using explicit constants from eq. (8), we get the final approximation form for the existence of the M\"ossbauer effect in magnetic field:

$$P \approx \frac{\pi}{2}\frac{\hbar^{3}c^{3}}{|Q|^{3}H^{3
}}\left(\frac{2|Q|H}{\hbar c} +  k^{2}\right)^{2}.\eqno(21)$$

Let us remark, that we can use approximation $e^{-2\alpha r^{2}} \approx 1 - 2\alpha r^{2}$.
Then instead of eq. (16), we write:

$$P \approx C^{2}\left|2\pi\int_{0}^{\infty}rdr\left[J_{0}(kr) - 2\alpha r^{2} J_{0}(kr)\right]\right|^{2}. \eqno(22)$$

Then using table integral (Gradshteyn and Ryzhik, 2007b)

$$\int_{0}^{\infty}x^{n}J_{l}(ax) dx = 2^{n}a^{-n-1}\frac{\Gamma\left(\frac{1}{2} + \frac{l}{2} + 
\frac{n}{2}\right)}{\Gamma\left(\frac{1}{2} + \frac{l}{2} - 
\frac{n}{2}\right)},\eqno(23)$$
we get:
$$P \approx C^{2}\left|\frac{2\pi}{k^{2}}\frac{\Gamma\left(1\right)}{\Gamma\left(0\right)} +
 \frac{32\pi\alpha}{k^{4}}\frac{\Gamma\left(2\right)}{\Gamma\left(-1\right)}\right|^{2}.\eqno(24)$$ 

To our surprise, this form of the  magnetic M\"ossbauer effect was not published
in the M\"ossbauer literature.

\section{Discussion}

The M\"ossbauer effect on magnetic field  is in no case the exact analogue of the M\"ossbauer effect in crystal, because magnetic field is the special physical reality (medium)  with unique quantum electrodynamic properties. 

In case that the decaying charged particle moves in accelerating potential $V = Fx$ where $F = - \partial V/\partial x$ is the accelerating force, then the corresponding Schr\"odinger equation in the momentum representation $(\hat x = i\hbar\partial/\partial p)$ is as follows (Drukarev, 1988):

$$\left(-i\hbar F \frac{\partial}{\partial p} + \frac{p^{2}}{2m} - E\right)\left<p|E\right> = 0\eqno(25)$$
with the solution 

$$\left<p|E\right> = \frac{1}{\sqrt{2\pi\hbar F}}\exp\left[\frac{i}{\hbar F}\left(Ep - \frac{p^{3}}{6m}\right)\right].\eqno(26)$$

Then, 

$$\left<x|E\right> = \int_{-\infty}^{\infty}\left<x|p\right>\left<p|E\right>dp = 
\frac{1}{2\pi\hbar F^{1/2}}\int_{-\infty}^{\infty}\exp\left\{
\frac{i}{\hbar}\left[\left(x + \frac{E}{F}\right)p -  \frac{p^{3}}{6mF}\right]\right\}dp,\eqno(27)$$
where we have used relation

$$\left<x|p\right> = \frac{1}{\sqrt{2\pi\hbar}}e^{ipx/\hbar}.\eqno(28)$$

The classical turning point is given by relation $V = E$, from which follows the coordinate of the turning point $x_{0} = E/F$. Then we write with regard to the last statement:

$$\left<x|E\right> =  
\frac{1}{2\pi\hbar F^{1/2}}\int_{-\infty}^{\infty}\exp\left\{
\frac{i}{\hbar}\left[(x - x_{0})p -  \frac{p^{3}}{6mF}\right]\right\}dp.\eqno(29)$$

After introducing the new variable

$$u = \frac{p}{(2m\hbar F)^{1/3}}; \quad z = \left(\frac{2mF}{\hbar^{2}}\right)^{1/2}(x_{0} - x), \eqno(30)$$
we get the solution of the Schr\"odinger equation for charged particle moving in the accelerated potential in the final form:

$$\left<x|E\right> =  
\left(\frac{2m}{\hbar^{2} F^{1/2}}\right)^{1/3}\frac{1}{2\pi}\int_{-\infty}^{\infty}\exp
\left[-i\left(\frac{u^{3}}{3} + zu\right)\right]dp,\eqno(31)$$
where 

$$v(z) = \frac{1}{2\pi}\int_{-\infty}^{\infty}\exp
\left[-i\left(\frac{u^{3}}{3} + zu\right)\right]dp\eqno(32)$$
is so called Airy function.

The final formula for the existence of the M\"ossbauer effect in the accelerated field is 

$$ P = \left|\int e^{i{\bf x}{\bf k}}|\left<x|E\right>|^{2}d{\bf x}\right|^{2}.\eqno(33)$$

 Ninio (1973) used instead of the Feynman amplitude the impulsive force $F(t) = const\,\delta(t-\lambda)$ to calculate the persistence of harmonic oscillator. After applying such impulsive force, the basic oscillator function is 

$$\psi_{0} = \exp\left(-\frac{1}{2}|\xi(t)|^{2}\right)
\sum_{m=0}^{\infty}\frac{[\xi(t)]^{2}}{(m!)^{1/2}}\phi_{m}, \eqno(34)$$
where 

$$\xi(t) = i(2m\hbar\omega)^{-1/2}Ae^{i\omega\lambda} \quad
(t>\lambda). \eqno(35)$$ 

The corresponding probability  of the basic state  persistence  is

$$P \approx |\left<\psi_{0}|\psi_{0}\right>|^{2} = e^{(-|\xi(t)|^{2})}
= e^{(-A^{2}/2m\hbar\omega)}. \eqno(36)$$

Thus, there is non zero probability that the impulse creates no phonons. However, it must be remembered that the oscillator particle is bound  to a fixed center. No doubt, that it is possible to use the Ninio method to calculate the M\"ossbauer effect in magnetic field and in the accelerated field. In addition, there is not excluded, that the so called maximal acceleration may play some role in case  of accelerated  charged particles. The recent discussion of the specific application  of the maximal acceleration in the M\"ossbauer physics was  presented by Potzel (2014). The  introduction of the maximal acceleration into physics by means of transformations between the reference systems was given by author (Pardy, 2003).  

The article is in a some sense the new mainstream  of ideas related to the M{\"o}ssbauer effect in physics and it can be applied in chemistry, biology, geology, cosmology,  medicine and other human activities. Let us remark that the discovery of the M{\"o}ssbauer was rewarded with the Nobel Prize in Physics in 1961 together with Robert Hofstadter's research of electron scattering in atomic nuclei. M\"ossbauer effect in the magnetic, or, electric field represents, the crucial problem for experimentalists and it is not excluded that the experimental realization of this effect leads to the adequate appreciation.

\vspace{7mm}
\noindent
{\bf REFERENCES}

\vspace{7mm}

\noindent
Berestetzkii, V. B.; Lifshitz, E. M. and Pitaevskii, L. P. (1989). {\it Quantum electrodynamics}, (Moscow, Nauka) (in Russian).\\[3mm]
Drukarev, G. F. (1988). {\it Quantum mechanics}, (Leningrad University Press), (in Russian). \\[3mm]
Feynman, R. (1972). {\it Statistical mechanics}, (W. B. Benjamin, Inc., Reading, Massachusetts).\\[3mm]
Gradshteyn, I. S. and Ryzhik, I. M. (2007a). {\it Tables of integrals, series and products}, Seven ed., (New York, Academic Press), section 3.461(3), page 364.\\[3mm]
Gradshteyn, I. S. and Ryzhik, I. M. (2007b). {\it Tables of integrals, series and products}, Seven ed., (New York, Academic Press), section 6.561(14), page 676.\\[3mm]
Kuznetsov, D. S. (1962). {\it The special functions}, Moscow. (in Russian). \\[3mm]
 M\"ossbauer. R. L. (1958). Kernresonanzfluoreszenz von Gammastrahlung
in $^{191}{\rm Ir}$, Z. Phys. {\bf 151}, 124. \\[3mm]
M\"ossbauer, R. L. (1958). Kernresonanzfluoreszenz von Gammastrahlung
in $^{191}{\rm Ir}$, Naturwissenschaften {\bf 45}, 538.\\[3mm]
M\"ossbauer, R. L. (1959). Kernresonanzabsorption von $\gamma$-Strahlung in $^{191}{\rm Ir}$, Z. Naturforsch. {\bf A 14}, 211.\\[3mm]
M\"ossbauer, R. L. (2000). The discovery of the M\"ossbauer effect,
Hyperfine Interactions {\bf 126}, 1.\\[3mm]
Ninio, F. (1973). The forced harmonic oscillator and the zero-phonon transition of the M\"ossbauer effect, AJP, {\bf 41}, 648-649.\\[3mm]
Pardy, M. (2003). The Space-time transformations between accelerated systems, arXiv:gr-qc/0302007, 
(gr-qc). \\[3mm]
Potzel, W. (2014). Clock hypothesis of relativity theory, maximal acceleration, and M\"ossbauer 
spectroscopy,  arXiv:1403.2412,  (physics.ins-det). \\[3mm] 
Rohlf, J. W. (1994). {\it Modern physics from $\alpha$ to $Z^{0}$},  (John Willey \& Sons, Inc., New York).
\end{document}